\documentclass{article}
\usepackage[affil-it]{authblk}
\usepackage[english]{babel}
\usepackage{amsmath, amsthm, amssymb}

\newcommand{\gen}[1]{{\left< #1 \right>}}
\newcommand{\ep}{\hfill $\blacksquare$}
\newcommand{\eep}{\hfill $\square$}
\newcommand{\pf}{\noindent {\bf Proof. \ }}

\newtheorem{theorem}{Theorem}[]
\newtheorem{lemma}[theorem]{Lemma}
\newtheorem{proposition}[theorem]{Proposition}
\newtheorem{corollary}[theorem]{Corollary}

\newtheorem{example}[theorem]{Example}
\newtheorem{remark}[theorem]{Remark}

\def\Z{{\mathbb Z}}
\def\F{{\mathbb F}}
\def\K{{\mathbb K}}
\def\R{{\mathcal R}}
\def\S{{\mathcal S}}
\def\M{{\mathcal M}}

\numberwithin{equation}{section}

\title{Left Dihedral Codes over Finite Chain Rings}

\author[1]{H. Aghili \thanks{Email: h.aghili@sci.ui.ac.ir}}
\author[2]{R. Sobhani \thanks{Corresponding author, Email: r.sobhani@sci.ui.ac.ir}}

\affil[1]{Department of Pure Mathematics, Faculty of Mathematics and Statistics, University of Isfahan,  Isfahan 81746-73441, Iran}
\affil[2]{Department of Applied Mathematics and Computer Science, Faculty of Mathematics and Statistics, University of Isfahan, Isfahan 81746-73441, Iran}

\date{}

\begin{document}
\maketitle

\begin{abstract}
Let $R$ be a finite commutative chain ring, $D_{2n}$ be the dihedral group of size $2n$ and $R[D_{2n}]$ be the dihedral group ring. In this paper, we completely characterize left ideals of $R[D_{2n}]$ (called left $D_{2n}$-codes) when ${\rm gcd}(char(R),n)=1$. In this way, we explore the structure of some skew-cyclic codes of length 2 over $R$ and also over $R\times S$, where $S$ is an isomorphic copy of $R$. As a particular result, we give the structure of cyclic codes of length 2 over $R$.  In the case where $R=\F_{p^m}$ is a Galois field, we give a classification for left $D_{2N}$-codes over $\F_{p^m}$, for any positive integer $N$. In both cases we  determine dual codes and identify self-dual ones.
\end{abstract}

{\bf Keywords:} Left dihedral codes; Chain rings; Skew-cyclic codes; Automorphism; Dual codes; Self-dual codes.

%**********************************************************************************************************************************************************************************************************************
%**********************************************************************************************************************************************************************************************************************
%**********************************************************************************************************************************************************************************************************************

\section{Introduction}

All rings in this paper are assumed to be finite commutative with identity except otherwise stated. For many years, Galois fields was the mostly used alphabet for classical codes until the authors of \cite{H,Nech} have discovered that many seemingly non-linear binary codes are in fact images of linear (extended cyclic) codes over the modular ring $\Z_4$. After that, many research works have been done on the structure of different families of codes over rings. From among different rings, the most attention has been paid to chain rings and from among linear codes, the most attention has been paid to cyclic, costacyclic and skew-cyclic codes.

Cyclic codes of length $n$ over a ring $R$ are in correspondence with ideals of the quotient ring $R_n:=R[x]/\gen{x^n-1}$. The ring $R_n$ is in fact the group ring $R[G]$ where $G=\gen{x}$ is a cyclic group of size $n$. Cyclic groups are of the most simplest form among finite groups. A bit more generalization of cyclic groups are abelian groups. Abelian groups are direct product of cyclic groups. When $G$ is an abelian group, the ideals of the group ring $R[G]$ are called abelian codes over $R$. Abelian codes over finite fields and finite chain rings have been studied by many researchers \cite{Bha,FM,JLLX,MR1,MR2,PM}.

For the general case when $G$ is an arbitrary finite group, $R[G]$ is non-commutative when $G$ is non-abelian and the structure of $R[G]$ becomes more complicated. Moreover, when $R[G]$ is non-commutative we have more objects to classify, namely left ideals. We say to a left ideal of $R[G]$, a left $G$-code. Generally, classification of left $G$-codes is a difficult task, but there are some research works dealing with the classification of two-sided $G$-codes \cite{BR,DFM,GI}. Also, for some special cases of finite non-abelian groups, there are research works dealing with the structure of left $G$-codes.

Let $D_{2n}=\gen{x,y\ |\ x^n=y^2=1,\ yx=x^{-1}y}$ be the dihedral group of size $2n$. Among the non-abelian groups, dihedral groups seems to be the most similar groups to cyclic ones. While the existence of a good family of cyclic codes is still an open problem, in \cite{BM}, it is shown that for infinitely many block lengths a random left $D_{2n}$-code is an asymptotically good rate-half code with a high probability. Also, in some papers \cite{ CCFG, M, MH}, it is shown that some  well-known good linear codes are left $D_{2n}$-codes. These results make left $D_{2n}$-codes interesting. In this field of research, the authors of \cite{B}, considered two-sided ideals of $\F_q[D_{2n}]$ when every prime factor of $n$ divides $q-1$. The authors of \cite{CCF} considered a more general case and classified all left ideals of $\F_q[D_{2n}]$ when ${\rm gcd}(n,q)=1$. This result was generalized to an special kind of chain rings in \cite{CCFW}, where the authors gave a characterization for left $D_{2n}$-codes over the Galois ring $GR(p^2,m)$, when ${\rm gcd}(p,n)=1$. In \cite{CCFG}, the authors gave a classification of binary left dihedral codes of length $8m$ and determined self-dual ones.

In the rest of this section, let $R$ be a chain ring of characteristic $p^k$. Let $n$ be a positive integer such that ${\rm gcd}(p,n)=1$.  In this paper, we completely classify left $D_{2n}$-codes over $R$. We then, for any positive integer $N$, classify all left $D_{2N}$-codes over the Galois field $\F_{p^m}$. We also determine dual codes and classify self-dual ones. These are in fact generalization to the results presented in \cite{CCF, CCFW, CCFG}.

As we will see, to complete our classification we need to characterize some skew-cyclic codes over $R$.  Let $\theta$ be an automorphism of $R$. A linear code $C$ of length $n$ over $R$ is said to be skew-cyclic code with respect to $\theta$ or briefly $\theta$-cyclic code if $$c=(c_0,c_1,\dots,c_{n-1})\in C\Longrightarrow (\theta(c_{n-1}),\theta(c_0),\dots,\theta(c_{n-2}))\in C.$$ The structure of skew-cyclic codes over fields is well-studied \cite{BGU,SAAS,BU} but fewer has been done for that over a chain ring \cite{BSU,JS}. Let $R[x;\theta]$ be the skew polynomial ring over $R$. In the case that the length $n$ is divisible by the order of $\theta$, the polynomial $x^n-1$ became a central element and skew-cyclic codes have an ideal structure. In fact they become left ideals of the ring $R[x;\theta]/\gen{x^n-1}$.  In this paper, when the order of $\theta$ divides 2, we characterize all skew-cyclic codes of length 2 over $R$, with respect to $\theta$. When $\theta$ is the identity automorphism, $\theta$-cyclic codes coincides with cyclic codes. The structure of cyclic codes of length $n$ over $R$, in the case where ${\rm gcd}(p,n)=1$, is well-studied \cite{DL}. The situation becomes more complicated when ${\rm gcd}(p,n)\ne 1$. Many works have been done in this case (\cite{Sa,DP,DLi,SE1,KLL,SE2}) but the problem is not settled yet. In particular, when $p=2$, the structure of cyclic codes of length $2$ over $R$ is open. Letting $\theta$ to be the identity map, we conclude the structure of such cyclic codes.

The structure of the paper is as follows. In the next section we give some preliminaries about chain rings and codes over chain rings. In Section \ref{DCCRS}, we give a general theorem on the structure of left $D_{2n}$-codes over $R$ for which ${\rm gcd}(p,n)=1$. We also state our main classification theorem in this section but the proof of the theorem can be completed using results of Sections \ref{SeSCCCR} and \ref{SeSCCDSCR}. More precisely, in Section \ref{SeSCCCR} we will explore the structure of some skew-cyclic codes of length 2 over $R$ and in Section \ref{SeSCCDSCR}, we will explore the structure of some skew-cyclic codes of length 2 over $R\times S$, where $S$ is isomorphic to $R$. In Section \ref{DC}, we determine dual codes and classify self-dual ones. Section \ref{DCFFS}, deals with the structure of left $D_{2N}$-codes and their duals over any Galois field $\F_{p^m}$, where $N$ is an arbitrary positive integer. The paper is closed with a conclusion section.

%**********************************************************************************************************************************************************************************************************************
%**********************************************************************************************************************************************************************************************************************
%**********************************************************************************************************************************************************************************************************************

\section{Preliminaries}\label{Sec 2}

In this section we present some necessary preliminaries. Recall that all rings in this paper are assumed to be finite, commutative and with identity. Let $R$ be a ring. $R$ is said to be {\it local} if it has only one maximal ideal, say $M$. $R$ is said to be a {\it chain} ring if its ideals form a chain under inclusion. It is well-known that $R$ is a chain ring if and only if it is a local ring whose maximal ideal $M$ is principally generated. Let $R$ be a  chain ring with $M=\gen{\gamma}$. Since elements of $M$ are nilpotent, so $\gamma$ is nilpotent. We denote the nilpotency index of $\gamma$, by $s$. Hence $\gamma^s=0$ and $\gamma^i\ne 0$ for $0\le i\le s-1$.

Since the quotient ring $R/M$ has no nontrivial ideal, it is a field, called the {\it residue field of $R$}. We denote by $\K$, the residue field of $R$. We also assume that $|\K|=q=p^m$, where $p$ is a prime and $m$, a positive integer. The field $\K$ may not be a subfield of $R$, but there is a well-known coset representatives of $M$ in $R$, called the {\it Trichm\"uller set}, and we denote it by $\tau_m$. More precisely, $R$ contains  a unit element $\zeta$ with multiplicative order $q-1$ for which $\tau_m=\{0,1,\zeta,\zeta^2,\dots,\zeta^{q-2}\}$. We call $\zeta$, the generator of $\tau_m$. Since the set $\tau_m$ modulo $\gamma$ equals $\K$, we do not make distinction between $\tau_m$ and $\K$. Any element $r$ in $R$ can be uniquely represented as $r=r_0+\gamma r_1+\cdots+\gamma^{s-1}r_{s-1}$, where $r_i\in\tau_m$. In this representation $r$ is unit if and only if $r_0\ne 0$. Using this representation, we define the {\it valuation} of $r$ to be the minimum integer $0\le i\le s-1$ for which $r_i\ne 0$.

It is known that the characteristic of $R$ is a power of the prime $p$, say $p^k$. In this case, the Galois ring $GR(p^k,m)$ is a subring of $R$. More precisely we have $R=GR(p^k,m)[x]/\gen{x^e+pf(x),x^s}$, where $1\le e\le s$ is a positive integer and $g(x):=x^e+pf(x)$ is a polynomial called the {\it Eisenstein polynomial}. From this representation for $R$, we can conclude that $p\in\gen{\gamma^e}$ and $p\notin\gen{\gamma^{e-1}}$. Consequently, $p=\epsilon\gamma^e$ for some unit element $\epsilon$ in $R$. Since $0=p^k=\epsilon^k\gamma^{ek}$ we have $\gamma^{ek}=0$ and hence $ek\ge s$. We write $s=e(k-1)+v$ where $1\le v\le e$ is a positive integer. Integers $p,k,m,s$ and $v$ are called {\it basic parameters} of $R$ and we will use these notation frequently in this paper. We also fix and use the notation $e$ and $\epsilon$ for $R$.

A code of length $n$ over $R$ is a subset of $R^n$. The code is said to be linear if it is a submodule of $R^n$. All codes in this paper are assumed to be linear.  Let $\mu$ be the map from $R$ to $\K$ which sends $r$ to $r$ modulo $\gamma$. The map $\mu$ can be naturally extended to $R^n$. Let $C$ be a code of length $n$ over $R$ and $0\le i\le s-1$. The $i$-th torsion code of $C$ over $\K$, denoted by ${\rm Tor}_i(C)$, is the code $${\rm Tor}_i(C)=\{\mu(c)\ |\ \gamma^ic\in C\}.$$ It is known that $|C|=\prod_{i=0}^{s-1}|{\rm Tor}_i(C)|$.

Now let $R'$ be another chain ring which includes $R$ and identities of $R$ and $R'$ are the same. In this case, $R'$ is said to be an extension of $R$. If the maximal ideal of $R'$ is $\M=\gen{\gamma'}$, then $R'$ is said to be a separable extension of $R$ if $\gamma=\gamma'$. Let $R'$ be a separable extention of $R$. An automorphism $f$ of $R'$  such that $f(r)=r$ for all $r\in R$, is called an $R$-automorphism of $R'$. The set of all $R$-automorphisms of $R'$ is called the Galois group of $R'$ over $R$ and is denoted by $G_R(R')$. It is known that $G_R(R')$ is a cyclic group (see \cite[Corollary XV.3]{Mc}) and we have $$G_R(R')\cong G_{R/M}(R'/\M).$$
In fact, there exists a primitive element $\omega$ of $R'$ over $R$ such that $\eta(\omega)=\omega^q$ is the generator of the cyclic group $G_R(R')$, where $q=|R/M|$ (see \cite[Theorem VX.10]{Mc}). Taking the generator of the Trichm\"uller set of $R'$ to be $\omega$, we see that for any $r'=r'_0+\gamma r'_1+\cdots+\gamma^{s-1}r'_{s-1}\in R'$ we have $$\eta(r')={r'_0}^q+\gamma {r'_1}^q+\cdots+\gamma^{s-1}{r_{s-1}}^q.$$ If $|R'/\M|=q^{2m'}$ then we call the $R$-automorphism $\eta^{m'}$, the Galois automorphism of order 2 of $R'$ over $R$.

%**********************************************************************************************************************************************************************************************************************
%**********************************************************************************************************************************************************************************************************************
%**********************************************************************************************************************************************************************************************************************

\section{Dihedral codes over chain rings}\label{DCCRS}

Let $G$ be a finite group of size $l$ and $R$ be a ring. Choose an order $g_1,g_2, \dots,g_l$ for the elements of $G$. The group ring $R[G]$ is a ring whose elements are all formal sums of the form $r_1g_1+r_2g_2+\cdots+r_lg_l$, where $r_i\in R$, for $1\le i\le l$. The addition and multiplication in $R[G]$ is done as
\begin{eqnarray*}
\left(\sum_{i=1}^{l}r_i g_i\right)+\left(\sum_{i=1}^{l}s_i g_i\right) & = & \left(\sum_{i=1}^{l}(r_i+s_i) g_i\right)\\
\left(\sum_{i=1}^{l}r_i g_i\right)\left(\sum_{i=1}^{l}s_i g_i\right) & = & \left(\sum_{i=1}^{l}c_i g_i\right),\\
\end{eqnarray*}
where $$c_i=\sum_{k,t: g_kg_t=g_i}r_ks_t.$$

Let $R$ be a ring, $R[y]$ be the ring of polynomials in $y$ with coefficients in $R$ and $\theta$ be an automorphism of $R$. We denote by $R[y;\theta]$, the ring of skew polynomials over $R$. Elements of $R[y;\theta]$ are precisely those of $R[y]$, addition in $R[y;\theta]$ is done similar to that in $R[y]$ and multiplication in $R[y;\theta]$ is done by the rule $ya=\theta(a)y$, for $a\in R$. Clearly, when $\theta$ is non-identity then $R[y;\theta]$ is non-commutative. It can be verified that if the order of $\theta$ divides $n$, then $y^n-1$ is a central element of $R[y;\theta]$, that is commute with all elements of $R[y;\theta]$. Consequently, the ideal generated by $y^n-1$ coincides with left multiples of $y^n-1$ and is a two-sided ideal. Therefore, the quotient ring $R_{n,\theta}:=R[y;\theta]/\gen{y^n-1}$ is meaningful. Left ideals of $R_{n,\theta}$ are known as skew-cyclic codes with respect to $\theta$ or $\theta$-cyclic codes of length $n$ over $R$.

The dihedral group of size $2n$, denoted by $D_{2n}$, is the group with the representation $D_{2n}=\gen{x,y\ |\ x^n=y^2=1,\ yx=x^{-1}y}$.

In the rest of this paper, let $R$ be a chain ring of characteristic $p^k$, maximal ideal $M=\gen{\gamma}$ and nilpotency $s$. We also let $n$ to be a positive integer for which ${\rm gcd}(p,n)=1$. In this section we present some general facts about the structure of the group ring $R[D_{2n}]$. We start with the following lemma. The proof can be simply derived from ordering $D_{2n}$ as $1,x,...,x^{n-1},y,xy,...,x^{n-1}y$ and hence we omit it.
\begin{lemma}\rm\label{GSL1}
The ring $R[D_{2n}]$ can be viewed as the ring $S[y;\theta]/\gen{y^2-1}$, where $S$ is the ring $R[x]/\gen{x^n-1}$ and $\theta$ is the automorphism of $S$ of order 2 which sends $f(x)$ to $f(x^{-1})$.\eep
\end{lemma}

Recall that for a polynomial $a(x)=a_0+a_1x+\cdots+a_dx^d$ in $R[x]$ with ${\rm deg}(a(x))=d$, we define the reciprocal of $a(x)$, denoted by $a^*(x)$, to be $$a^*(x):=x^da(1/x).$$ Now, since ${\rm gcd}(p,n)=1$, the polynomial $x^n-1$ factors uniquely as the product of irreducible polynomials over $R$. Write $$x^n-1=f_1(x)f_2(x)...f_r(x)f_{r+1}(x)...f_{r+w}(x),$$ where,
\begin{itemize}
\item for $1\le i\le r$, $f_i(x)$ is irreducible and $f_i^*(x)=f_i(x)$,
\item for $r+1\le i\le r+w$, $f_i(x)=g_i(x)g_i^*(x)$ where $g_i(x)$ is irreducible with $g_i^*(x)\ne g_i(x)$.
\end{itemize}
Also, we have $$S=\bigoplus_{i=1}^{r+w}A_i,$$ where $A_i:=\gen{\widehat{f_i(x)}}$ and $\widehat{f_i(x)}=(x^n-1)/f_i(x)$. Now, for $1\le i\le r$, the ideal $A_i$ is a chain ring of size $|R|^{{\rm deg}(f_i(x))}$ with an idempotent generator $E_i(x)$ which is its identity. The idempotent $E_i(x)$ can be obtained easily. Since ${\rm gcd}(f_i(x),\widehat{f_i(x)})=1$, there are polynomials $a_i(x), b_i(x)$ such that $$a_i(x)\widehat{f_i(x)}+b_i(x)f_i(x)=1.$$ Now it can be seen that $E_i(x)=a_i(x)\widehat{f_i(x)}$. Another fact about $A_i$ is that the maximal ideal of $A_i$ is $M_i=\gen{\gamma E_i(x)}$ and its nilpotency is that for $R$, namely $s$. Furthermore, for $r+1\le i\le r+w$ we have $A_i=A_{i_1}\oplus A_{i_2}$ where $A_{i_1}=\gen{\widehat{g_i(x)}}$ and $A_{i_2}=\gen{\widehat{g_i^*(x)}}$. Both $A_{i_1}$ and $A_{i_2}$ are chain rings of size $|R|^{{\rm deg}(g_i(x))}$ and with idempotent generators $E_{i_1}(x)$ and $E_{i_2}(x)$ as their identities, respectively (these idempotents can be obtained similarly to that obtained for $A_i$, $1\le i\le r$). Note that the identity of $A_i$ is $E_i(x)=E_{i_1}(x)+E_{i_2}(x)$.

Now, it can be easily seen that the restriction of $\theta$ to each component ring $A_i$, denoted by $\theta_i$, is an automorphism of $A_i$. More precisely, for $1\le i\le r$, the ring $A_i$ is a separable extension of $R\gen{E_i(x)}$ and $\theta_i$ is the identity automorphism of $A_i$ (when ${\rm deg}(f_i(x))=1$) or an automorphism of $A_i$ of order 2 which fixes $R\gen{E_i(x)}$. Hence $\theta_i$ is identity or the Galois automorphism of order 2 of $A_i$ over $R\gen{E_i(x)}$. Note that when $f_i^*(x)=f_i(x)$ and ${\rm deg}(f_i(x))\ge 2$ then ${\rm deg}(f_i(x))$ must be an even number. Also, for $r+1\le i\le r+w$, $\theta_i$ is the automorphism of $A_i=A_{i_1}\oplus A_{i_2}$ which sends $a+b$ to $\varphi^{-1}(b)+\varphi(a)$, where $\varphi: A_{i_1}\longrightarrow A_{i_2}$ is a ring isomorphism which sends $a(x)$ to $a(x^{-1})$. Now we have proven the following theorem.
\begin{theorem}\label{GT}\rm
With notation as above, we have
\begin{eqnarray*}
R[D_{2n}] & = & \frac{S[y;\theta]}{\gen{y^2-1}}\\
               & = & \frac{(\bigoplus_{i=1}^{r+w}A_i)[y;\theta]}{\gen{y^2-1}}\\
               & = & \bigoplus_{i=1}^{r}\frac{A_i[y;\theta_i]}{\gen{E_i(x)y^2-E_i(x)}}\bigoplus_{i=r+1}^{r+w}\frac{A_i[y;\theta_i]}{\gen{E_i(x)y^2-E_i(x)}},
\end{eqnarray*}
where for $1\le i\le r$, either $\theta_i$ is the identity automorphism of $A_i$ (when ${\rm deg}(f_i(x))=1$) or the Galois automorphism of order 2 of $A_i$ over $R\gen{E_i(x)}$, and for $r+1\le i\le r+w$, $\theta_i$ is the automorphism of $A_i=A_{i_1}\oplus A_{i_2}$ which sends $a+b$ to $\varphi^{-1}(b)+\varphi(a)$ and $\varphi: A_{i_1}\longrightarrow A_{i_2}$ is the ring isomorphism which sends $a(x)$ to $a(x^{-1})$.
\eep
\end{theorem}
In Section \ref{SeSCCCR}, we will completely explore the structure of left ideals of the rings $A_i[y;\theta_i]/\gen{E_i(x)y^2-E_i(x)}$ when $1\le i\le r$ and in Section \ref{SeSCCDSCR} we will do that for $r+1\le i\le r+w$. In fact, we first give the structure of $\theta_i$-cyclic codes of length 2 over$A_i$, $1\le i\le r$, in Section \ref{SeSCCCR}, and then give the structure of $\theta_i$-cyclic codes of length 2 over $A_i=A_{i_1}\oplus A_{i_2}$, $r+1\le i\le r+w$, in Section \ref{SeSCCDSCR}. Using results of these two sections, we can conclude the following theorem which is the main structural theorem on dihedral codes over chain rings.%\newpage
\begin{theorem}\label{MT}\rm
Let $C$ be a left $D_{2n}$-code over $R$, that is a left ideal of $R[D_{2n}]$. Then we have $C=\bigoplus_{i=1}^{r+w}C_i$, where
\begin{itemize}
\item[A)] for $1\le i\le r$, $C_i$ is of one of the following forms:
\begin{itemize}
\item[A1)] $C_i=\gen{\gamma^{b_i}E_i(x)}$ where $0\le b_i\le s$. Also we have $$|C_i|=q^{2{\rm deg}(f_i(x))(s-b_i)}.$$
\item[A2)] $C_i=\gen{\gamma^{a_i}E_i(x)(y+r_i(x)),\gamma^{b_i}E_i(x)}$, where $0\le a_i<b_i\le s$ and $r_i(x)$ is a unit in $A_i$ for which $r_i(x)r_i(x^{-1})=E_i(x)\bmod\gamma^{b_i-a_i}$. Also we have $$|C_i|=q^{{\rm deg}(f_i(x))(2s-a_i-b_i)},$$ and the polynomials $r_i(x)$ can be obtained using Propositions \ref{WtProp} and \ref{WtProp2} in Section \ref{SeSCCCR}.
\end{itemize}
\item[B)] for $r+1\le i\le r+w$, $C_i$ is of one of the following forms:
\begin{itemize}
\item[B1)] $C_i=\gen{\gamma^{a_i}E_{i_1}(x)+\gamma^{c_i}E_{i_2}(x)}$, where $0\le a_i, c_i\le s$. Also we have $$|C_i|=q^{2{\rm deg}(g_i(x))(2s-a_i-c_i)}.$$
\item[B2)] $C_i=\gen{(\gamma^{a_i}E_{i_1}(x)+\gamma^{c_i}E_{i_2}(x))+(\gamma^{b_i}\alpha_i(x))y}$,  where $0\le a_i\le s-1$, $1\le c_i\le s$, ${\rm max}\{0,a_i+c_i-s\}\le b_i\le c_i-1$ and $\alpha_i(x)$ is a unit in $A_{i_1}\bmod\gamma^{c_i-b_i}$. Also we have $$|C_i|=q^{2{\rm deg}(g_i(x))(2s-a_i-c_i)}.$$
\end{itemize}
\end{itemize}
\end{theorem}
\pf The proof follows from Theorems \ref{MT}, \ref{MatIso} and \ref{LCL2}. \ep

%**********************************************************************************************************************************************************************************************************************
%**********************************************************************************************************************************************************************************************************************
%**********************************************************************************************************************************************************************************************************************

\section{$\theta_i$-cyclic codes of length 2 over $A_i$ for $1\le i\le r$}\label{SeSCCCR}

As we mentioned, in this section we explore the structure of $\theta_i$-cyclic codes of length 2 over the chain rings $A_i$, $1\le i\le r$. To this end, for simplicity, we represent the chain ring $A_i$ with $\R$, the generator of the maximal ideal of $A_i$, namely $\gamma E_i(x)$,  with $\Gamma$ and the identity of $A_i$, namely $E_i(x)$, with $1$. We also consider a more general case and instead of $\theta_i$, we use an arbitrary automorphism $\Theta$ of $\R$ for which ${\rm ord}(\Theta)\in\{1,2\}$.

We use $s$ for the nilpotency of $\Gamma$ and $d$ for the degree of $f_i(x)$. Note that the residue field of $\R$ is $\F_{q^{d}}$. For $1\le t\le s$, let $\R_t$ be the chain ring $\R/\gen{\Gamma^t}$. Even though, in general, $\R_t$ is not a subring of $\R$, we may consider it as a subset of $\R$. Set $$W_{t,\Theta}:=\{r\in\R_t\ |\ r\Theta(r)=1\bmod\Gamma^t\}.$$ Let us denote the quotient ring $\R[y;\Theta]/\gen{y^2-1}$ by $\S$.  In the next theorem we give a complete classification for left ideals of $\S$ which is based on the sets $W_{t,\Theta}$.
\begin{theorem}\label{MT}\rm
Any left ideal $I$ of $\S$ is of one of the following unique forms:
\begin{itemize}
\item[(1)] $I=\gen{\Gamma^b}$ for some $0\le b\le s$.
\item[(2)] $I=\gen{\Gamma^a(y+r),\Gamma^b}$ where $0\le a<b\le s$ and $r\in W_{b-a,\Theta}$.
\end{itemize}
Moreover, in case $(1)$ we have $|C|=q^{2d(s-b)}$ and in case $(2)$ we have $$|C|=q^{d(2s-(a+b))}.$$
\end{theorem}
\pf Let $I$ be an ideal of $\S$. Let $a$ be the minimum valuation of leading coefficients of elements of degree one in $I$, and $b$ be the minimum valuation of elements of degree zero (constants) in $I$. Hence there exist polynomials $f_1(y)=\Gamma^ay+\Gamma^c\alpha$ and $g(y)=\Gamma^b$ in $I$ such that $\alpha\in\R$ is zero or a unit. Since $\Theta$ is an automorphism of $\R$, we have $\Theta(\Gamma)=\beta\Gamma$ for some unit $\beta\in\R$. Now since $\beta^{-b}y\Gamma^b=\Gamma^b y\in I$, we have $a\le b$. First, we show that $I=\gen{f_1(y),g(y)}$. To see this, assume that $h(y)=\Gamma^{u_1}\alpha_1y+\Gamma^{u_2}\alpha_2$ is an arbitrary element of $I$. If $\alpha_1=0$ or $\alpha_2=0$ then clearly we have $h(y)\in\gen{g(y)}$. Assume that both of $\alpha_1$ and $\alpha_2$ are units. Since $u_1\ge a$ we have $$f_2(y):=\Gamma^{u_1-a}\alpha_1f_1(y)-h(y)=\Gamma^{u_2}\alpha_2-\Gamma^{u_1-a+c}\alpha\alpha_1\in I.$$ Since $f_2(y)$ is constant, there exist $\alpha_3$ in $\R$ and non-negative integer $u_3$ such that $f_2(y)=\Gamma^{u_3}\alpha_3.$ Consequently $u_3\ge b$ and hence $f_2(y)=\Gamma^{u_3-b}\alpha_3g(y)$. Therefore
$$h(y)=\Gamma^{u_1-a}\alpha_1f_1(y)-\Gamma^{u_3-b}\alpha_3g(y)\in \gen{f_1(y),g(y)}.$$

Now if $a=b$ then easily we can see that $f_1(y)\in\gen{\Gamma^b}$ and hence $I=\gen{\Gamma^b}$. If $a<b$ then we claim that $a$ must be equal to $c$. To see this, first note that if $a>c$ then
\begin{eqnarray*}
\Theta(\alpha)^{-1}\beta^{-c}yf_1(y) & = & \Gamma^cy+\Gamma^a\beta^{a-c}\Theta(\alpha)^{-1}\in I,
\end{eqnarray*}
which is a contradiction with the choice of $a$. Now if $a<c$ then since we have
\begin{eqnarray*}
\Gamma^{c-a}f_1(y) & = & \Gamma^cy+\Gamma^{2c-a}\alpha\in I,
\end{eqnarray*}
we conclude that
\begin{eqnarray}\label{EQ1}
\Theta(\alpha)^{-1}\beta^{-c}yf_1(y)-\Gamma^{c-a}f_1(y)= \Gamma^a(\beta^{a-c}\Theta(\alpha)^{-1}-\Gamma^{2(c-a)}\alpha)\in I.
\end{eqnarray}
Since $c>a$, $\beta^{a-c}\Theta(\alpha)^{-1}-\Gamma^{2(c-a)}\alpha$ is a unit and hence we deduce that $\Gamma^a\in I$ which is a contradiction with the choice of $b$ and the fact that $a<b$. Therefore we must have $a=c$ and $f_1(y)=\Gamma^a(y+\alpha)$. Now if $\alpha=0$ then we conclude that $\Gamma^a\in I$ which is a contradiction. Hence $\alpha$ is a unit. Moreover, from Equation $(\ref{EQ1})$, we conclude that
\begin{eqnarray*}
\Gamma^a(\Theta(\alpha)^{-1}-\alpha)=0\bmod \Gamma^{b-a},
\end{eqnarray*}
or equivalently $\alpha\Theta(\alpha)=1 \bmod \Gamma^{b-a}$. Write $\alpha=\alpha_0+\alpha_1\Gamma+\cdots+\alpha_{s-1}\Gamma^{s-1}$. Set $r:=\alpha_0+\alpha_1\Gamma+\cdots+\alpha_{b-a-1}\Gamma^{b-a-1}$ and $f(y):=\Gamma^a(y+r)$. Clearly,
$r\in W_{b-a,\Theta}$, $f(y)\in I$ and $I=\gen{f(y),g(y)}$. For uniqueness, note that clearly $g(y)$ is unique. If $F(y)=\Gamma^a(y+r')$ be an alternative for $f(y)$ then we have $$f(y)-F(y)=\Gamma^a(r-r').$$ Hence $\Gamma^{b-a}$ must divide $r-r'$ and therefore
$r=r'$.

On the other hand, similar arguments show that if we consider $$I=\gen{\Gamma^a(y+r),\Gamma^b},$$ where $0\le a<b\le s$ and $r\in W_{b-a,\Theta}$ then $a$ is the minimum valuation of leading coefficients of elements of degree one in $I$, and $b$  is the minimum valuation of elements of degree zero (constants) in $I$. So such an $I$ is in the unique form. Also ideals of the form $I=\gen{\Gamma^b}$ where $0\le b\le s$ are in the unique form. Consequently all distinct ideals of $\S$ are precisely those listed in parts $(1)$ and $(2)$ of the
theorem.

For the cardinality assertion note that $|I|=\prod_{i=0}^{s-1}|{\rm Tor}_i(I)|$. Now, in case $(1)$ we have $|{\rm Tor}_i(I)|=1$ for $0\le i\le b-1$ and $|{\rm Tor}_i(I)|=q^{2d}$ for $b\le i\le s-1$. Also in case $(2)$ we have $|{\rm Tor}_i(I)|=1$ for $0\le i\le a-1$, $|{\rm Tor}_i(I)|=q^d$ for $a\le i\le b-1$ and $|{\rm Tor}_i(I)|=q^{2d}$ for $b\le i\le s-1$. The proof is now completed.\ep

To complete the classification, we need to determine the sets $W_{t,\Theta}$. In the next of the section, we completely determine these sets when $\Theta$ is the identity or the unique Galois automorphism of $\R$ of order 2. Let $m'=md$ and $\tau_{m'}$ be the Trichm\"uller set of $\R$ and $\tau_{m'}^*$ denotes the unit elements in $\tau_{m'}$. Since $W_{t,\Theta}$ is subgroup of $\R^*$ and $\R^*\cong\tau_{m'}^*\times(1+\Gamma\R)$ we may write $W_{t,\Theta}\cong W^{(1)}_{t,\Theta}\times W^{(2)}_{t,\Theta}$ where $W^{(1)}_{t,\Theta}$ is a subgroup of $\tau_{m'}^*$ and $W^{(2)}_{t,\Theta}$ is a subgroup of $1+\Gamma\R$.

Now, let $id$ denotes the identity automorphism of $\R$. In the next proposition, we will obtain $W_{t,id}$ for $1\le t\le s$ and calculate $|W_{t,id}|$. Recall that $p=\epsilon\Gamma^e$ form some unit $\epsilon\in\R$ and some positive integer $e$.
\begin{proposition}\rm\label{WtProp}
If $q$ is odd then for each $1\le t\le s$ we have $W^{(1)}_{t,id}=\{1,-1\}$
and $W^{(2)}_{t,id}=\{1\}$. Hence $|W_{t,id}|=2$. If $q$ is even then we have
\begin{itemize}
\item[(I)] If $1\le t\le 2e$ then $W^{(1)}_{t,id}=\{1\}$ and
$W^{(2)}_{t,id}=\{1+\Gamma^{\lceil t/2\rceil}r\ |\ r\in\R_{\lfloor t/2\rfloor}\}$.
Hence $|W_{t,id}|=q^{d\lfloor t/2\rfloor}$.
\item[(II)] If $2e+1\le t\le s$ then $W^{(1)}_{t,id}=\{1\}$ and
$W^{(2)}_{t,id}=\{\pm1+\Gamma^{t-e}r\ |\ r\in\R_e\}$. Hence $|W_{t,id}|=2q^{de}$.
\end{itemize}
\end{proposition}
\pf
Clearly, $W^{(1)}_{t,id}$ is $\{1, -1\}$ when $q$ is odd and $\{1\}$ when $q$ is even. We hence focus on $W^{(2)}_{t,id}$. Let $r\in W^{(2)}_{t,id}$. Write $r=1+\Gamma^l r_0$ where $r_0$ is zero or a unit. We have
\begin{eqnarray*}
1 & = & r^2\\
  & = & 1+\Gamma^lr_0(2+\Gamma^l r_0).
\end{eqnarray*}
Hence $\Gamma^lr_0(2+\Gamma^l r_0)=0$. If $q$ is odd then $2+\Gamma^l r_0$ is a unit and hence we must have $\Gamma^lr_0=0$. Therefore, $1$ is the only solution for $r$. Now let $q$ be even. Since $2=\epsilon\Gamma^e$, where $\epsilon$ is a unit in $\R$ and $e$ is a positive integer, we have
\begin{eqnarray*}
0 & = & \Gamma^lr_0(2+\Gamma^l r_0)\\
  & = & r_0(\epsilon\Gamma^{l+e}+\Gamma^{2l}r_0).
\end{eqnarray*}
Hence we have $r_0=0$ or, $r_0\ne 0$ and
\begin{equation}\label{EQ2}
\epsilon\Gamma^{l+e}+\Gamma^{2l}r_0=0.
\end{equation}
Now we have four possibilities:
\begin{itemize}
\item[(A)] ${\rm min}\{l+e,2l\}\ge t$.
\item[(B)] $l+e\ge t$ and $2l<t$.
\item[(C)] $l+e<t$ and $2l\ge t$.
\item[(D)] ${\rm max}\{l+e,2l\}< t$.
\end{itemize}
Clearly cases $(B), (C)$ lead to contradiction. Also if $(A)$ holds then clearly $(\ref{EQ2})$ holds. But $(A)$ implies that $l\ge t-e$ and also $l\ge t/2$. Now if $t-e>t/2$ or equivalently $e<t/2$ then $l$ varies between $t-e$ and $t-1$. Also if $e\ge t/2$ then $l$ varies between $\lceil t/2\rceil$ and $t-1$.  Consequently case $(I)$ and a part of case $(II)$ are covered. Finally, we discuss on $(D)$. In this case, if $l+e\ne 2l$ then again we lead to a contradiction from $(\ref{EQ2})$. Hence $l+e=2l$ or equivalently $l=e$. Therefore Equation $(\ref{EQ2})$ becomes $\Gamma^{2e}(\epsilon+r_0)=0$. Hence $\Gamma^{t-2e}\mid (\epsilon+r_0)$ and thus $r_0=-\epsilon+r_1\Gamma^{t-2e}$. Consequently $r=1+\Gamma^er_0=1+\Gamma^e(-\epsilon+\Gamma^{t-2e}r_1)$ and hence $r=-1+\Gamma^{t-e}r_1$. This completes the proof.
\ep

Now we are able to conclude the structure of cyclic codes of length 2 over chain rings. We show a chain ring with $R$ and use all notation previously fixed for that.
\begin{corollary}\rm\label{CCCRC}
Let $C$ be a cyclic code of length 2 over $R$, that is an ideal of the ring $R[y]/\gen{y^2-1}$. If $q$ is odd then $C$ is of one of the following unique forms:
\begin{itemize}
\item[1)] $C=\gen{\gamma^b}$ for some $0\le b\le s$.
\item[2)] $C=\gen{\gamma^a(y\pm 1),\gamma^b}$ where $0\le a<b\le s$.
\end{itemize}
Also, if $q$ is even then $C$ is of one of the following unique forms:
\begin{itemize}
\item[1)] $C=\gen{\gamma^b}$ for some $0\le b\le s$.
\item[2)] $C=\gen{\gamma^a(y+r),\gamma^b}$ where $0\le a<b\le s$, $1\le b-a\le 2e$ and $$r\in\{1+\gamma^{\lceil (b-a)/2\rceil}r'\ |\ r'\in R_{\lfloor (b-a)/2\rfloor}\}.$$
\item[3)] $C=\gen{\gamma^a(y+r),\gamma^b}$ where $0\le a<b\le s$, $b-a>2e$ and $$r\in\{\pm1+\gamma^{b-a-e}r'\ |\ r'\in R_e\}.$$
\end{itemize}\eep
\end{corollary}

Now, let $m'=2l$ be an even integer and $\Theta$ be the order 2 Galois automorphism of $\R$.  Let $\tau_{m'}=\{0,1,\zeta,\zeta^2,\dots,\zeta^{q^d-2}\}$, where $\zeta$ is an element of order $q^d-1$ in $\R$. In the next proposition, we will obtain $W_{t,\Theta}$ for all $1\le t\le s$.
\begin{proposition}\rm\label{WtProp2}
Let $A$ be the subgroup $\gen{\zeta^{p^l-1}}$ then for all $1\le t \le s$ we have $W^{(1)}_{t,\Theta}=A$. Also, if $t=1$ then $W^{(2)}_{t,\Theta}=\{1\}$. Moreover for $2\le t\le s$ we have
\begin{itemize}
\item[(I)] if $q$ is even then for each $r\in W^{(2)}_{t-1,\Theta}$ there exist a coset $D_r$ of the additive subgroup $\F_{p^l}$ in $\F_q$ for which we have $$W^{(2)}_{t,\theta}=\{r+\Gamma^{t-1}\delta\ |\ r\in W^{(2)}_{t-1,\Theta}, \delta\in D_r\}.$$
\item[(II)] if $q$ is odd then for each $r\in W^{(2)}_{t-1,\Theta}$ there exist a coset $D_r$ of the additive subgroup $\{0\}\cup\zeta^{2(p^l-1)}\F_{p^l}^*$ in $\F_q$ for which we have $$W^{(2)}_{t,\Theta}=\{r+\Gamma^{t-1}\delta\ |\ r\in W^{(2)}_{t-1,\Theta}, \delta\in D_r\}.$$
\end{itemize}
Consequently we have $|W_{t,\Theta}|=(p^{l}+1)p^{l(t-1)}$, for all $1\le t\le s$.
\end{proposition}
\pf Clearly we have $W^{(1)}_{t,\Theta}=A$. Let $a\in W^{(2)}_{t,\Theta}$ and write $a=r+\Gamma^{t-1}\delta$ where $r\in (1+\Gamma R_{t-2})$ and $\delta\in\tau_m$. Noting that $r=1+\Gamma r'$ for some $r'$, we have
\begin{eqnarray*}
1 & = & a\Theta(a)\\
  & = & r\Theta(r)+\Gamma^{t-1}(\delta+\Theta(\delta)).
\end{eqnarray*}
Hence we need to have $r\Theta(r)=1\bmod\Gamma^{t-1}$ and therefore, there exist $\beta_r\in\tau_m$ such that $r\Theta(r)=1+\Gamma^{t-1}\beta_r$. Consequently we have
\begin{eqnarray*}
1 & = & 1+\Gamma^{t-1}(\delta+\Theta(\delta)+\beta_r).
\end{eqnarray*}
Thus we need to have $\delta+\Theta(\delta)+\beta_r=0\bmod\Gamma$. Let us define $f:\F_q\rightarrow\F_{p^l}$ by $f(\alpha)=\alpha+\Theta(\alpha)$. Clearly $f$ is an additive homomorphism between two additive groups. Now, we need to find all solutions of the equation $f(\delta)=-\beta_r$, for a given $\beta_r$. But when $q$ is even, the kernel of $f$ is $K=\F_{p^l}$ and thus the solutions are all $\delta$ in the coset $D_r$ of the additive subgroup $\F_{p^l}$ of $\F_q$ for which $f$ takes the value $-\beta_r$ on that coset. The solutions can be obtained for the case $q$ is odd, by noting that the kernel of $f$ is $K=\{0\}\cup\zeta^{2(p^l-1)}\F_{p^l}^*$. The proof is now completed. \ep

\begin{remark}\label{IO2}\rm
As we mentioned, we consider a general automorphism of $\R$ which is of order 2 or the identity map. But in our study we need only two possibilities for $\Theta$. Since for $1\le i\le r$, $f_i(x)$ is a self-reciprocal polynomial, the degree of $f_i(x)$, $d$, is equal to one or is an even integer.
When $d=1$, the automorphism $\theta_i$ becomes the identity automorphism of $A_i$ and when $d\ne 1$, it must be an even integer and $\Theta$ becomes the unique Galois automorphism of order 2 of $\R$ over $R\gen{E_i(x)}$. Hence Propositions \ref{WtProp} and \ref{WtProp2} suffices for completely determining
$\theta_i$-cyclic codes of length 2 over $A_i$, when $1\le i\le r$.
\end{remark}
\begin{remark}\label{OA}\rm
As we can see from Propositions \ref{WtProp} and \ref{WtProp2}, cardinality of $W_{t,\Theta}$ strongly depends on $\Theta$. Also its of worth mentioning that, there are other kinds of automorphisms of order 2 for a finite chain ring, rather than the Galois one. For example, let $p$ be an odd prime, $l$ be a positive integer and $R$ be the chain ring $\F_{p^{2l}}[u]/\gen{u^2}$. Let $\Theta$ be the automorphism of $R$ which sends $a+ub$ to $a-ub$. Clearly this automorphism is not the Galois automorphism. Also, easily one can check that $|W_{1,\Theta}|=2$. Let $a+bu$ be in $W_{2,\Theta}$. Hence we must have
\begin{eqnarray*}
1 & = & (a+bu)(a-bu)\\
  & = & a^2.
\end{eqnarray*}
Therefore we conclude that $a\in\{1,-1\}$ and $b\in\F_{p^{2l}}$ which proves that $|W_{2,\Theta}|=2p^{2l}$.
\end{remark}

%**********************************************************************************************************************************************************************************************************************
%**********************************************************************************************************************************************************************************************************************
%**********************************************************************************************************************************************************************************************************************

\section{$\theta_i$-cyclic codes of length 2 over $A_i$ for $r+1\le i\le r+w$}\label{SeSCCDSCR}

In this section we explore the structure of $\theta_i$-cyclic codes of length 2 over the rings $A_i$, $r+1\le i\le r+w$. To this end, we consider a bit more general case. Since for $r+1\le i\le r+w$, we have $A_i=A_{i_1}\oplus A_{i_2}$ where $A_{i_1}$ and $A_{i_2}$ are two isomorphic chain rings, for simplicity, we represent the chain ring $A_{i_1}$ with $\R$ and $A_{i_2}$ with $\R'$, the generator of the maximal ideal of $A_i$, namely $\gamma E_{i_1}(x)$,  with $\Gamma$, the size of the residue field of $\R$ with $q^d=p^{md}$, $d={\rm deg}(g_i(x))$, and the nilpotency of $\Gamma$ by $s$. We also use the external direct product of $\R$ and $\R'$, denoted by $\R\times\R'$, instead of the direct sum. Recall that, for $r+1\le i\le r+w$, $\theta_i$ is the automorphism of $A_i=A_{i_1}\oplus A_{i_2}$ which sends $a+b$ to $\varphi^{-1}(b)+\varphi(a)$, where $\varphi: A_{i_1}\longrightarrow A_{i_2}$ is a ring isomorphism which sends $a(x)$ to $a(x^{-1})$. Here, we consider a bit more general case and let $\Theta$ to be the automorphism of $\R\times\R'$ which sends $(r,r')$ to $(\varphi^{-1}(r'),\varphi(r))$ where $\varphi:\R\longrightarrow\R'$ is any ring isomorphism.

Set $T:=\R\times\R'$. Clearly $\Theta$ is an automorphism of $T$ of order 2. Let us show the identity of $T$, namely $(E_{i_1}(x),E_{i_2}(x))$, by $1$. In what follows in this section, we classify left ideals of the quotient ring $T[y;\Theta]/\gen{y^2-1}$, that is $\Theta$-cyclic codes of length 2 over $T$. First we show in the following theorem that the ring $T[y;\Theta]/\gen{y^2-1}$ is isomorphic to the ring of $2\times 2$ matrices over $\R$, namely $M_2(\R)$. This is a key result in our classification, since then we characterize left ideals of the ring $M_2(\R)$, by using the Morita Equivalence and Theorem \ref{LCL2} which deals with the classification of linear codes of length 2 over a finite chain ring.
\begin{theorem}\label{MatIso}\rm
Two rings  $T[y;\Theta]/\gen{y^2-1}$ and $M_2(\R)$ are isomorphic.
\end{theorem}
\pf
Define the map $\psi: T[y;\Theta]/\gen{y^2-1}\longrightarrow M_2(\R)$ by
\begin{eqnarray*}
\psi\left((r_1,r'_1)+(r_2,r'_2)y\right)=
\left(
\begin{array}{cc}
r_1 & r_2\\
\varphi^{-1}(r'_2) & \varphi^{-1}(r'_1)
\end{array}
\right).
\end{eqnarray*}
It can be easily seen that $\psi$ is a ring isomorphism and we are done.
\ep

Two rings $A$ and $B$ are said to be Morita equivalence if there is a one to one correspondence between the set of $A$-modules and the set of $B$-modules. It is known that, for a chain ring $A$, $A$ and $M_n(A)$ are Morita equivalence. For our case, the chain ring $\R$ is Morita equivalence with $M_2(\R)$. More precisely, left ideals of $M_2(\R)$ are in correspondence with $\R$-modules of $\R\times\R$, that is linear codes of length 2 over $\R$. In fact, the generator matrix of a linear code of length 2 over $\R$ can be considered as the generator of the corresponding left ideal of $M_2(\R)$. In order to complete our classification, we need to characterize all distinct linear codes of length 2 over the chain ring $\R$. We do this in the following theorem.
\begin{theorem}\label{LCL2}\rm
The generator matrix $G$ of any linear code $C$ of length 2 over the chain ring $\R$ is of one of the following forms.
\begin{itemize}
\item[1)] Type I:
\[G=\left(
\begin{array}{cc}
\Gamma^a & 0\\
0 & \Gamma^c
\end{array}
\right),\] where $0\le a,c\le s$.
\item[2)] Type II:
\[G=\left(
\begin{array}{cc}
\Gamma^a & \Gamma^b\alpha\\
0 & \Gamma^c
\end{array}
\right),\] where $0\le a\le s-1$, $1\le c\le s$, ${\rm max}\{0,a+c-s\}\le b\le c-1$ and $\alpha$ is a unit in $\R\bmod \Gamma^{c-b}$.
\end{itemize}
Moreover, in each case we have $|C|=q^{d(2s-a-c)}$.
\end{theorem}
\pf
For $1\le i\le 2$, let $\pi_i:C\longrightarrow\R$ be the projection map on $i$-th component. Let $I^{(1)}$ be $\pi_1(C)$. Since $I^{(1)}$ is an ideal of $\R$, there exists $0\le a\le s$ such that $I^{(1)}=\gen{\Gamma^a}$. Also there exists a codeword $c^{(1)}=(\Gamma^a,\Gamma^b\alpha)$ in $C$ for which $\alpha$ is zero or a unit. Now let $I^{(2)}:=\{\pi_2(v)\ |\ v\in C, \pi_1(v)=0\}$. Again $I^{(2)}$ is an ideal of $\R$ and hence there exists integer $c$ such that$ I^{(2)}=\gen{\Gamma^c}$. Also there exists a codeword $c^{(2)}=(0,\Gamma^c)$ in $C$. It is easy to verify that the set $\{c^{(1)},c^{(2)}\}$ generates $C$. If $\alpha\ne 0$ we also may assume that $\alpha$ is a unit in $\R\bmod\Gamma^{c-b}$. Again, easily it can be proved that such a set of generators of $C$ is unique. Therefore, $C$ has the unique generator matrix of the form
\[G=\left(
\begin{array}{cc}
\Gamma^a & \Gamma^b\alpha\\
0 & \Gamma^c
\end{array}
\right),\] where $I^{(1)}=\gen{\Gamma^a}$, $I^{(2)}=\gen{\Gamma^c}$ and $\alpha$ is zero or a unit in $\R\bmod\Gamma^{c-b}$. Moreover, since $(0,\Gamma^{s-a+b})$ lies in $C$ we need to have $s-a+b\ge c$. Conversely, one can see that, for any code $C$ with a generator matrix of the form given above and satisfying the given conditions on $a,b,c,\alpha$, we have $I^{(1)}=\gen{\Gamma^a}$, $I^{(2)}=\gen{\Gamma^c}$ and $C$ is in the unique form. Separating two cases $\alpha=0$ and $\alpha\ne 0$ we get the result. For the cardinality assertion, note that for $0\le i< c$, we have ${\rm Tor}_i(C)=\{0\}$ and for $c\le i\le a$ we have ${\rm Tor}_i(C)$ is generated by the single vector $(0,1)$ and finally, for $a+1\le i\le s$ we have ${\rm Tor}_i(C)$ is a whole two-dimensional space $\F_{q^d}\times\F_{q^d}$. Hence  we have $|C|=q^{d(a-c)}q^{2d(s-a)}=q^{d(2s-a-c)}$ as desired. The proof is now completed.
\ep
\begin{remark}\rm
If $C$ is a linear code of length 2 over $\R$, then its corresponding ideal $I$ in $M_2(\R)$ is formed from all matrices whose rows are codewords of $C$. Hence we have $|\psi^{-1}(I)|=|I|=|C|^2$.
\end{remark}

%**********************************************************************************************************************************************************************************************************************
%**********************************************************************************************************************************************************************************************************************
%**********************************************************************************************************************************************************************************************************************

\section{Dual codes}\label{DC}

In this section, we first investigate the duals of the codes described in Theorem \ref{MT} and then will give a classification for self-dual codes. Let
\[\begin{array}{l}
a=\left(a_{1,1}, a_{1,2}, \ldots, a_{1, n}, a_{2,1}, \ldots, a_{2, n}\right),\\
b=\left(b_{1,1}, b_{1,2}, \ldots, b_{1, n}, b_{2,1}, \ldots, b_{2, n}\right)
\end{array}\]
be two elements of $R^{2n}$. For $i=1,2$, we also set $a_{i}(x):=\sum_{j=1}^{n} a_{i, j} x^{j-1}$ and $b_{i}(x):=\sum_{j=1}^{n} b_{i, j} x^{j-1}$. The Euclidean inner product of $a$ and $b$ is defined to be $\sum_{j=1}^{n} \sum_{i=1}^{2} a_{i, j} b_{i, j}$ and is denoted by $[a, b]$. For a linear code $C$ of length $2n$ over $R$, we define the Euclidean dual of $C$ to be $\left\{b \in R^{2 n} \mid[a, b]=0, \forall a \in C\right\}$ and denote it by $C^{\perp}$. We say that $C$ is self-dual if $C=C^{\perp}$.
\begin{lemma}\rm\label{ssaa}
With notation as above, if $$\left(a_{1}(x)+a_{2}(x) y\right)\left(b_{1}\left(x^{-1}\right)+b_{2}(x) y\right)=0$$ in $\S[x]/\gen{x^n-1}$, then we have $[a, b]=0$. Also, if we set $I:=<a_{1}(x)+a_{2}(x)y>$ and  $J:=<b_{1}(x)+b_{2}(x) y>$, then we have $J \subseteq I^{\perp}$.
\end{lemma}
\pf
\begin{flalign*}
&(a_{1}(x)+a_{2}(x) y)\left(b_{1}(x^{-1})+b_{2}(x) y\right) \\ &=a_{1}(x) b_{1}(x^{-1})+a_{2}(x) b_{2}(x^{-1})+(a_{1}(x) b_{2}(x)+a_{2}(x) b_{1}(x)) y \\ &=\left(\sum_{j=1}^{n} \sum_{i=1}^{2} a_{i, j} b_{i, j}+g_{1} x+\cdots+g_{n-1} x^{n-1}\right)+h(x) y\ \ \bmod (x^{n}-1)
\end{flalign*}
where $h(x) \in S$ and for $1 \leq t \leq n-1$ we have $g_{t} \in R$. Therefore if $$\left(a_{1}(x)+a_{2}(x) y\right)\left(b_{1}\left(x^{-1}\right)+b_{2}(x) y\right)=0,$$ then we must have $[a, b]=0$. Now, let $$a(x)+b(x) y=\left(a'_{1}(x)+a'_{2}(x)y\right)\left(a_{1}(x)+a_{2}(x) y\right)\in I$$ and $c(x)+d(x)y=\left(b'_{1}(x)+b'_{2}(x)y\right)\left(b_{1}(x)+b_{2}(x)y\right)\in J$. We have
\[\begin{array}{ll}
	\left(a(x)+b(x)y\right)\left(c\left(x^{-1}\right)+d(x) y\right) =  \\
 	\left(a_{1}^{\prime}(x)+a_{2}^{\prime}(x) y\right)\left(a_{1}(x)+a_{2}(x) y\right)\left(b_{1}\left(x^{-1}\right) b_{1}^{\prime}\left(x^{-1}\right)+b_{2}(x) b_{2}^{\prime}\left(x^{-1}\right) +  \right.\\
    \left.\left(b_{1}^{\prime}(x) b_{2}(x)+b_{2}^{\prime}(x) b_{1}\left(x^{-1}\right)\right) y\right)  =  \\
 	\left(a_{1}^{\prime}(x)+a_{2}^{\prime}(x) y\right) \underbrace{\left(a_{1}(x)+a_{2}(x) y\right)\left(b_{1}\left(x^{-1}\right)+b_{2}(x) y\right)}_{=0}\left(b_{1}^{\prime}\left(x^{-1}\right)+b_{2}^{\prime}(x) y\right)  =  0.\\
\end{array}\]
This implies that $J\subseteq I^{\perp}$ and the proof is completed.\ep\\

\noindent The following corollary follows easily from Lemma \ref{ssaa} and hence we omit the proof.
\begin{corollary}\rm\label{aabb}
Let $I=<a_{1}(x)+a_{2}(x) y, a_{3}(x)>$ and $J=<b_{1}(x)+b_{2}(x) y, b_{3}(x)>$. If all four conditions below hold then we have $J \subseteq I^{\perp}$.
\begin{itemize}
\item[(1)]  $\left(a_{1}(x)+a_{2}(x) y\right)\left(b_{1}\left(x^{-1}\right)+b_{2}(x) y\right)=0$.
\item[(2)] $a_{3}(x)\left(b_{1}\left(x^{-1}\right)+b_{2}(x) y\right)=0$.
\item[(3)] $\left(a_{1}(x)+a_{2}(x) y\right) b_{3}\left(x^{-1}\right)=0$.
\item[(4)] $a_{3}(x) b_{3}\left(x^{-1}\right)=0$.\eep
\end{itemize}
\end{corollary}
Now we are able to determine duals of the codes described in Theorem \ref{MT}. We do this in the following theorem.
\begin{theorem}\label{SS}\rm
Let $C=\bigoplus_{i=1}^{r+w}C_i$ be a left $D_{2n}$-code over $R$. Then we have $C^{\perp}=\bigoplus_{i=1}^{r+w}D_i$, where
\begin{itemize}
\item[A)] for $1\le i\le r$, $D_i$ is of one of the following forms:
\begin{itemize}
\item[A1)] If $C_i=\gen{\gamma^{b_i}E_i(x)}$ then $D_i=\gen{\gamma^{s-b_i}E_i(x)}$. Also we have $$|D_i|=q^{2{\rm deg}(f_i(x))b_i}.$$
\item[A2)] If $C_i=\gen{\gamma^{a_i}E_i(x)(y+r_i(x)),\gamma^{b_i}E_i(x)}$ then $$D_i=\gen{\gamma^{s-b_i}E_i(x)(y-r_i(x)),\gamma^{s-a_i}E_i(x)}.$$ Also we have $|D_i|=q^{{\rm deg}(f_i(x))(a_i+b_i)}$.
\end{itemize}
\item[B)] For $r+1\le i\le r+w$, $D_i$ is of one of the following forms:
\begin{itemize}
\item[B1)] If $C_i=\gen{\gamma^{a_i}E_{i_1}(x)+\gamma^{c_i}E_{i_2}(x)}$ then $D_i=\gen{\gamma^{s-c_i}E_{i_1}(x)+\gamma^{s-a_i}E_{i_2}(x)}$. Also we have $|D_i|=q^{2{\rm deg}(g_i(x))(a_i+c_i)}$.
\item[B2)] If $C_i=\gen{(\gamma^{a_i}E_{i_1}(x)+\gamma^{c_i}E_{i_2}(x))+(\gamma^{b_i}\alpha_i(x)y}$ then $$D_i=\gen{(\gamma^{s-c_i}E_{i_1}(x)+\gamma^{s-a_i}E_{i_2}(x))+(\gamma^{s+b_i-a_i-c_i}(-\alpha_i(x))y}.$$ Also we have $|D_i|=q^{2{\rm deg}(g_i(x))(a_i+c_i)}$.
\end{itemize}
\end{itemize}
\end{theorem}
\pf
First note that any $D_i$ is in its unique form described in Theorem \ref{MT}. Moreover we have
\begin{flalign*}
|C||D|=\prod_{i=1}^{r} |C_i||D_i| \prod_{i=r+1}^{r+w} |C_i||D_i|=q^{2 s\left(\sum_{i=1}^{r} \operatorname{deg}\left(f_{i}\right)(x)+2 \sum_{i=1}^{r} \operatorname{deg}\left(g_{i}\right)(x)\right)}=|R|^{2 n}
\end{flalign*}
and hence $|D|=|C^{\perp}|$. Now by Lemma \ref{ssaa} and Corollary \ref{aabb}, one can easily verify that $D_{i} \subseteq C_{i}^{\perp}$ for every $1\le i\le r+w$. Therefore $$D=\bigoplus_{i=1}^{r+w} D_{i} \subseteq \bigoplus_{i=1}^{r+w} C_{i}^{\perp} \subseteq C^{\perp}.$$ Consequently $D=C^{\perp}$ and the proof is completed.\ep\\

\noindent We now classify self-dual codes in the following corollary.
\begin{corollary}\rm\label{ccbb}
Let $C=\bigoplus_{i=1}^{r+w}C_i$ be a self-dual left $D_{2n}$-code over $R$.
We have the following possibilities:
\begin{itemize}
\item
If $char(R)=2$ and $s$ is even, then
\begin{itemize}
\item[A)] for $1\le i\le r$, $C_i$ is of one of the following types:
\begin{itemize}
\item[A1)] Type I:  $C_i=\gen{\gamma^{s/2}E_i(x)}$.
\item[A2)] Type II: $C_i=\gen{\gamma^{a_i}E_i(x)(y+r_i(x)),\gamma^{s-a_i}E_i(x)}$, where $0\le a_i< s/2$ and $r_i(x)\in A_i$ is such that $$r_i(x)r_i(x^{-1})=E_i(x)\bmod\gamma^{s-2a_i}.$$
\end{itemize}
\item[B)] for $r+1\le i\le r+w$, $C_i$ is of one of the following types:
\begin{itemize}
\item[B1)] Type III: $C_i=\gen{\gamma^{a_i}E_{i_1}(x)+\gamma^{s-a_i}E_{i_2}(x)}$, where $0\le a_i\le s$.
\item[B2)] Type IV: $C_i=\gen{(\gamma^{a_i}E_{i_1}(x)+\gamma^{s-a_i}E_{i_2}(x))+(\gamma^{b_i}\alpha_i(x))y}$,  where $0\le a_i\le s-1$, $0\le b_i\le s-a_i-1$ and $\alpha_i(x)$ is a unit in $A_{i_1}\bmod\gamma^{s-a_i-b_i}$.
\end{itemize}
\end{itemize}
\item If $char(R)=2$ and $s$ is odd, then for $1\le i\le r$, $C_i$ is of Type II and for $r+1\le i\le r+w$, $C_i$ is of Type III or IV.
\item If $char(R)\neq2$ and $s$ is even, then for $1\le i\le r$, $C_i$ is of Type I and for $r+1\le i\le r+w$, $C_i$ is of Type III.
\item If $char(R)\neq2$ and $s$ is odd, then there is no self-dual code in this case.
\end{itemize}
\eep
\end{corollary}

%**********************************************************************************************************************************************************************************************************************
%**********************************************************************************************************************************************************************************************************************
%**********************************************************************************************************************************************************************************************************************

\section{Dihedral codes over finite fields}\label{DCFFS}

In this section we will explore the structure of left ideals of $\F_{p^m}D_{2N}$, where $N=p^ln$ with ${\rm gcd}(n,p)=1$. We also determine dual codes and classify self-dual ones. Note that we can write $$x^N-1=f_1(x)^{p^l}f_2(x)^{p^l}...f_r(x)^{p^l}f_{r+1}(x)^{p^l}...f_{r+w}(x)^{p^l},$$ where,
\begin{itemize}
\item for $1\le i\le r$, $f_i(x)$ is irreducible and $f_i^*(x)=f_i(x)$,
\item for $r+1\le i\le r+w$, $f_i(x)=g_i(x)g_i^*(x)$ where $g_i(x)$ is irreducible with $g_i^*(x)\ne g_i(x)$.
\end{itemize}
Also we have the following lemma.
\begin{lemma}\rm\label{GSL2}
The ring $\F_{p^m}D_{2N}$ can be viewed as the ring $S[y;\theta]/\gen{y^2-1}$, where $S$ is the ring $\F_{p^m}[x]/\gen{x^N-1}$ and $\theta$ is the automorphism of $S$ of order 2 which sends $f(x)$ to $f(x^{-1})$.\eep
\end{lemma}
We have $S=\bigoplus_{i=1}^{r+s}A_i$ where $A_i:=\gen{(x^N-1)/f_i(x)^{p^l}}$. Now, for $1\le i\le r$, the ideal $A_i$ is a chain ring of size $p^{mp^l{\rm deg}(f_i(x))}$ with an idempotent generator $E_i(x)$ which is its identity. The maximal ideal of $A_i$ is $M_i=\gen{ E_i(x)f_i(x)}$ and its nilpotency is $p^l$. Moreover, for $r+1\le i\le r+w$ we have $A_i=A_{i_1}\oplus A_{i_2}$ where $A_{i_1}=\gen{(x^N-1)/g_i(x)^{p^l}}$ and $A_{i_2}=\gen{(x^N-1)/g_i^*(x)^{p^l}}$. Both $A_{i_1}$ and $A_{i_2}$ are chain rings of size $p^{mp^l{\rm deg}(g_i(x))}$ and with idempotent generators $E_{i_1}(x)$ and $E_{i_2}(x)$ as their identities (and hence the identity of $A_i$ is $E_i(x)=E_{i_1}(x)+E_{i_2}(x)$), respectively. Now, it can be easily seen that the restriction of $\theta$ to each component ring $A_i$, denoted by $\theta_i$, is an automorphism of $A_i$. Moreover, for $r+1\le i\le r+w$, $\theta_i$ is the automorphism of $A_i=A_{i_1}\oplus A_{i_2}$ which sends $a+b$ to $\varphi^{-1}(b)+\varphi(a)$, where $\varphi: A_{i_1}\longrightarrow A_{i_2}$ is a ring isomorphism which sends $a(x)$ to $a(x^{-1})$. Now we have proven the following theorem.
\begin{theorem}\label{GT}\rm
With notation as above, we have
\begin{eqnarray*}
\F_{p^m}D_{2N} & = & \frac{S[y;\theta]}{\gen{y^2-1}}\\
               & = & \frac{(\bigoplus_{i=1}^{r+w}A_i)[y;\theta]}{\gen{y^2-1}}\\
               & = & \bigoplus_{i=1}^{r}\frac{A_i[y;\theta_i]}{\gen{E_i(x)y^2-E_i(x)}}\bigoplus_{i=r+1}^{r+w}\frac{A_i[y;\theta_i]}{\gen{E_i(x)y^2-E_i(x)}},
\end{eqnarray*}
where for $1\le i\le r$ we have  $\theta_i$ is the automorphism of $A_i$ which sends $a(x)$ to $a(x^{-1})$, and for $r+1\le i\le r+w$ we have $\theta_i$ is the automorphism of $A_i=A_{i_1}\oplus A_{i_2}$ which sends $a+b$ to $\varphi^{-1}(b)+\varphi(a)$ and $\varphi: A_{i_1}\longrightarrow A_{i_2}$ is the ring isomorphism which sends $a(x)$ to $a(x^{-1})$.
\eep
\end{theorem}
\begin{theorem}\label{MTGF}\rm
Let $C$ be a dihedral code of length $2N$ over $\F_{p^m}$, that is a left ideal of $\F_{p^m}D_{2N}$. Then we have $C=\bigoplus_{i=1}^{r+w}C_i$, where
\begin{itemize}
\item[A)] for $1\le i\le r$, $C_i$ is of one of the following forms:
\begin{itemize}
\item[A1)] $C_i=\gen{(E_i(x)f_i(x))^{b_i}}$ for some $0\le b_i\le p^l$. Also we have $$|C_i|=p^{2m{\rm deg}(f_i(x))(p^l-b_i)}.$$
\item[A2)] $C_i=\gen{(E_i(x)f_i(x))^{a_i}(y+r_i(x)),(E_i(x)f_i(x))^{b_i}}$, where $$0\le a_i<b_i\le p^l$$ and $r_i(x)\in A_i$ is such that $r_i(x)r_i(x^{-1})=E_i(x)\bmod(E_i(x)f_i(x))^{b_i-a_i}$. Also we have $$|C_i|=p^{m{\rm deg}(f_i(x))(2p^l-(a_i+b_i))}.$$
\end{itemize}
\item[B)] for $r+1\le i\le r+w$, $C_i$ is of one of the following forms:
\begin{itemize}
\item[B1)] $C_i=\gen{(E_{i_1}(x)g_i^*(x))^{a_i}+(E_{i_2}(x)g_i(x))^{c_i}}$, where $0\le a_i, c_i\le s$. Also we have $$|C_i|=p^{2m{\rm deg}(g_i(x))(2p^l-a_i-c_i)}.$$
\item[B2)] $C_i=\gen{((E_{i_1}(x)g_i^*(x))^{a_i}+(E_{i_2}(x)g_i(x))^{c_i})+((E_{i_1}(x)g_i^*(x))^{b_i}\alpha_i(x))y}$,  where $0\le a_i\le s-1$, $1\le c_i\le s$, ${\rm max}\{0,a_i+c_i-s\}\le b_i\le c_i-1$ and $\alpha_i(x)$ is a unit in $A_{i_1}\bmod(E_{i_1}(x)g_i(x))^{c_i-b_i}$. Also we have $$|C_i|=p^{2m{\rm deg}(g_i(x))(2p^l-a_i-c_i)}.$$
\end{itemize}
\end{itemize}
\end{theorem}
\pf The proof follows from Theorems \ref{MT}, \ref{MatIso}, \ref{LCL2} and \ref{GT}. \ep\\
\begin{remark}\rm\label{R3}
For $1\le i\le r$, the authomorphism of $A_i$ is not the Galois automorphism. Also it is not the identity map even if ${\rm deg}(f_i(x))=1$. For example, Let $N=4$, $p=2$ and $m=1$. In this case we have $x^N-1=x^4+1=(x+1)^4$ and $A_1=\gen{1}=S=\F_2[x]/\gen{x^4+1}$. Also $\theta_1:A_1\longrightarrow A_1$ sends $a(x)$ to $a(x^{-1})$ and hence we have $\theta_1(x)=x^3$. This shows that $\theta_1$ is not the identity map. Moreover, $\theta_1$ does not fix the generator of the maximal ideal of $A_1$, namely $x+1$ since $\theta_1(x+1)=x^3+1$. As a consequence, Propositions \ref{WtProp} and \ref{WtProp2} can not be used to determine polynomials $r_i(x)$ in sub-case $A2$ of Theorem \ref{MTGF}.
\end{remark}
\noindent We now determine dual codes in the following theorem.
\begin{theorem}\label{SSF}\rm
Let $C=\bigoplus_{i=1}^{r+w}C_i$ be a left $D_{2N}$-code over $\F_{p^m}$. Then we have $C^{\perp}=\bigoplus_{i=1}^{r+w}D_i$, where
\begin{itemize}
\item[A)] for $1\le i\le r$, $D_i$ is of one of the following forms:
\begin{itemize}
\item[A1)] If $C_i=\gen{(E_i(x)f_i(x))^{b_i}}$ then $D_i=\gen{(E_i(x)f_i(x))^{p^l-b_i}}$. Also we have $|D_i|=p^{2m{\rm deg}(f_i(x))b_i}.$
\item[A2)] If $C_i=\gen{(E_i(x)f_i(x))^{a_i}(y+r_i(x)),(E_i(x)f_i(x))^{b_i}}$ then $$D_i=\gen{(E_i(x)f_i(x))^{p^l-b_i}(y-r_i(x)),(E_i(x)f_i(x))^{p^l-a_i}}.$$ Also we have $|D_i|=p^{m{\rm deg}(f_i(x))(a_i+b_i)}$.
\end{itemize}
\item[B)] For $r+1\le i\le r+w$, $D_i$ is of one of the following forms:
\begin{itemize}
\item[B1)] If $C_i=\gen{(E_{i_1}(x)g_i^*(x))^{a_i}+(E_{i_2}(x)g_i(x))^{c_i}}$ then $$D_i=\gen{(E_{i_1}(x)g_i^*(x))^{p^l-c_i}+(E_{i_2}(x)g_i(x))^{p^l-a_i}}.$$ Also we have $|D_i|=p^{2m{\rm deg}(g_i(x))(a_i+c_i)}$.
\item[B2)] If $C_i=\gen{((E_{i_1}(x)g_i^*(x))^{a_i}+(E_{i_2}(x)g_i(x))^{c_i})+((E_{i_1}(x)g_i^*(x))^{b_i}\alpha_i(x))y}$ then {\scriptsize{ $$ D_i=\gen{((E_{i_1}(x)g_i^*(x))^{p^l-c_i}+(E_{i_2}(x)g_i(x))^{p^l-a_i})+((E_{i_1}(x)g_i^*(x))^{p^l+b_i-a_i-c_i}\alpha_i(x))y}.$$}  }
    Also we have $|D_i|=p^{2m{\rm deg}(g_i(x))(a_i+c_i)}$.
\end{itemize}
\end{itemize}
\end{theorem}
\pf The proof follows from Theorems \ref{SS} and \ref{MTGF}. \ep

\noindent The following corollary classifies self-dual codes. For the proof it suffices to note that $s=p^l$ and $p$ is the characteristic of $S$. In fact $s$ is even if and only if $p=2$.
\begin{corollary}\rm\label{ccbbF}
Let $C=\bigoplus_{i=1}^{r+w}C_i$ be a self-dual left $D_{2N}$-code over $\F_{p^m}$.
We have the following possibilities:
\begin{itemize}
\item
If $p=2$, then
\begin{itemize}
\item[A)] for $1\le i\le r$, $C_i$ is of one of the following types:
\begin{itemize}
\item[A1)] Type I:  $C_i=\gen{(E_i(x)f_i(x))^{2^{l-1}}}$.
\item[A2)] Type II: $C_i=\gen{(E_i(x)f_i(x))^{a_i}(y+r_i(x)),(E_i(x)f_i(x))^{2^l-a_i}}$,\\ where $0\le a_i< 2^{l-1}$ and $r_i(x)\in A_i$ is such that $$r_i(x)r_i(x^{-1})=E_i(x)\bmod\gamma^{2^l-2a_i}.$$
\end{itemize}
\item[B)] for $r+1\le i\le r+w$, $C_i$ is of one of the following types:
\begin{itemize}
\item[B1)] Type III: $C_i=\gen{(E_{i_1}(x)g_i^*(x))^{a_i}+(E_{i_2}(x)g_i(x))^{2^l-a_i}}$, where $0\le a_i\le 2^l$.
\item[B2)] Type IV:  { \scriptsize{$$C_i=\gen{((E_{i_1}(x)g_i^*(x))^{a_i}+(E_{i_2}(x)g_i(x))^{2^l-a_i})+((E_{i_1}(x)g_i^*(x))^{b_i}\alpha_i(x))y},$$}} where $0\le a_i\le 2^l-1$, $0\le b_i\le 2^l-a_i-1$ and $\alpha_i(x)$ is a unit in $A_{i_1}\bmod (E_{i_1}(x)g_i^*(x))^{2^l-a_i-b_i}$.
\end{itemize}
\end{itemize}
\item If $p\ne 2$, then there is no self-dual code in this case.
\end{itemize}
\eep
\end{corollary}
\begin{example}\rm
In this example we restate that the doubly-even self-dual binary codes with parameters $(24,12,8)$ (the Golay code) and $(48,24,12)$, are binary left $D_{2N}$-codes that are principally generated. Note that these results previously obtained in \cite{M,MH}. Since $x^3+1=(x+1)(x^2+x+1)$ and both $x+1$ and $x^2+x+1$ are self-reciprocal, we are in the case $A$ of Theorem \ref{GT} and it suffices to determine the polynomials $r_1(x), r_2(x)$ with integers $a_1, a_2,b_1,b_2$. For the binary Golay code, we set $b_1=b_2=4$ (hence the code is principally generated) and $$r_1(x)=x^{10}+x^8+x^5+x^4+x^2+x$$ and $$r_2(x)=x^{11}+x^{10}+x^9+x^7+x^6+x^5+x^3+x^2+x.$$ Easily one can see that the generator matrix of the corresponding code (as a quasi-cyclic code) is $[1,r_1(x)E_1(x)+r_2(x)E_2(x)]$, where $E_1(x)=x^8+x^4$ and $E_2(x)=x^8+x^4+1$. Also, for the binary code with parameters $(48,24,12)$, we set $b_1=b_2=8$ (again the code is principally generated) and $$r_1(x)=x^{23}+x^{21}+x^{17}+x^{16}+x^{10}+x^9+x^8+x^7+x^5+x^2$$ and $$r_2(x)=x^{22}+x^{20}+x^{16}+x^{14}+x^{12}+x^8+x^6+x^4+1.$$ Again, the generator matrix of this code in its quasi-cyclic form is $$[1,r_1(x)E_1(x)+r_2(x)E_2(x)],$$ where $E_1(x)=x^{16}+x^8$ and $E_2(x)=x^{16}+x^8+1$.
\end{example}

%**********************************************************************************************************************************************************************************************************************
%**********************************************************************************************************************************************************************************************************************
%**********************************************************************************************************************************************************************************************************************

\section{Conclusion and remarks}

The complete structure of dihedral codes over finite chain rings with characteristic coprime with the half of the length, was explored. As a conclusion, the structure of dihedral codes of arbitrary length over finite fields was given. To derive a full characterization in this case, one need to compute elements $r$ in a chain ring $R$ with maximal ideal generated by $\gamma$ of nilpotency $s$ such that $r\theta(r)=1\bmod\gamma^t$, where $\theta$ is an special automorphism of $R$ and $1\le t\le s$. This seems to be a challenging open problem in a general form such that $\theta$ is an arbitrary automorphism of $R$. A variation of this problem has been introduced in \cite{CCFG,SZWZ}.

%**********************************************************************************************************************************************************************************************************************
%**********************************************************************************************************************************************************************************************************************
%**********************************************************************************************************************************************************************************************************************

\end{document}